# Unveiling the Oxidation Mechanisms of Octa-Penta Graphene: A Multidimensional Exploration from First-Principles to Machine Learning

Chenyi Zhou,[1] Rubin Huo,[1] Boyi Situ,[1] Zihan Yan,[2] Zhe Zhang,[1,*] Yusong Tu,[1]

[1]*College of Physics Science and Technology, Yangzhou University, Jiangsu 225009, China*
[2]*School of Engineering, Westlake University, Hangzhou, Zhejiang 300024, China*
*Corresponding Authors: zzhang@yzu.edu.cn

Abstract

Octa-penta graphene (OPG), a novel carbon allotrope characterized by its distinctive arrangement of pentagonal and octagonal rings, has garnered considerable attention due to its exceptional structure and functional properties. This study systematically investigates the oxidation mechanisms of OPG and elucidates the oxygen migration patterns on the OPG monolayer through first-principles calculations and machine-learning-based molecular dynamics (MLMD) simulations. Specifically, the oxidation processes on OPG-*L* and OPG-*Z* involve exothermic chemisorption, where oxygen molecules dissociate at the surfaces, forming stable epoxy groups. Notably, OPG-*Z* requires higher initial activation energy, reflecting the variable energy demands across different surfaces. The most energetically favorable adsorption site for an oxygen atom on OPG-*L* and OPG-*Z* are the $L_{8\text{-}5\text{-}2}$ site and the $Z_{8\text{-}5\text{-}1}$ site respectively, confirmed by their low adsorption energies and optimal bond configurations. Furthermore, the integrated-crystal orbital Hamilton population (ICOHP) and Bader charge analyses provide insights into the physical mechanisms of oxygen atom adsorption. Importantly, we found that oxidation also impact the electronic properties of OPG, with OPG-*L* retaining its metallic characteristics post-oxygen adsorption, whereas OPG-*Z* undergoes a transformation from a metallic to a semiconducting state due to the introduction of oxygen. Oxygen migration on OPG monolayer involves breaking and reforming of C-O bonds, with varying stability across adsorption sites and limited migration along the basal plane. MLMD simulations corroborate these migration patterns, offering detailed migration trajectories consistent with theoretical predictions. These findings enhance the understanding of oxygen migration dynamics on OPG, facilitate its experimental validations, and highlight its potential as a novel 2D material for applications in batteries, heat-resistant materials, and oxidation-resistant coatings.

## 1. Introduction

Octa-penta graphene (OPG) is a novel graphene derivative composed of octagonal and pentagonal carbon ring. Compared to traditional hexagonal graphene, OPG has a unique lattice structure that imparts distinct electronic, optical, and mechanical properties. Its high mechanical strength, good electrical conductivity, and excellent thermal conductivity make it promising for applications in molecular electronics, nanoelectronics and energy storage[1-3]. For instance, theoretical studies suggest that OPG exhibits significant thermal anisotropy due to its unique topology featuring non-hexagonal rings. This anisotropy can be controlled through strain, temperature, and defect engineering, making it crucial for effective thermal management in nano-electromechanical systems[4]. Additionally, the distinctive structure of OPG has been identified as a promising high-performance anode material for lithium-ion batteries, providing high theoretical capacity and efficient lithium adsorption and diffusion[5]. However, due to its complex lattice structure, OPG still faces several challenges in practical applications, particularly in the regulation and optimization of material properties.

Oxidation is a key method for the functionalization of two-dimensional (2D) materials, including graphene and its derivatives[6-10]. Introducing oxygen atoms can significantly alter the electronic properties and surface chemistry of materials, enhancing their performance in various applications, such as nanoelectronics[11-15], energy storage and conversion[16-20], and protective coatings[21-22]. Oxidation can modulate the bandgap of graphene, increase its chemical activity, and improve its interface interactions with other materials[8, 10, 23-27]. However, oxidation can also damage the atomic thin structures of 2D materials, introduce structural defects, and irreversibly alter their electronic, optical, and chemical characteristics[28-30]. For instance, black phosphorus, a promising material for electronic devices, suffer from poor chemical stability due to its sensitivity to oxidation[31-32]. Similarly, in the field of energy storage, MXene experiences a significantly reduction in performance effectiveness upon oxidation, affecting its application in electrochemical energy storage[30, 32]. Therefore, in-depth research on the oxidation process of OPG is crucial for optimizing its current applications and developing new functionalities. Such research is essential for evaluating the stability of OPG in oxygen-rich environments and assessing its feasibility as an oxidation-resistant coating material.

In this work, we employ first-principles density functional theory calculations and machine-learning-based molecular dynamics (MLMD) simulations to systematically elucidate the oxidation mechanisms of OPG and the dynamic oxygen migration patterns on the OPG monolayer. By calculating the energy, electronic density of states, and band structures of OPG under different oxidation states, we aim to reveal the adsorption behavior of oxygen atoms on OPG and their regulatory mechanisms on the material's electronic properties. Additionally, we also essentially delineate the physical mechanisms underlying the stable adsorption of oxygen atoms on OPG, including the crystal orbital Hamilton populations (COHP), the charge density difference and Bader charges. Our analysis reveals significant variations in the energy barriers governing oxygen migration across different chemisorption sites on the OPG monolayer. These findings offer essential insights into the fundamental oxidation processes on the OPG surface and the dynamic behaviors of oxygen atoms on this distinctive octa-penta topological structure.

## 2. Computational model and methods

*2.1 Model of OPG monolayer*

Different morphological combinations of pentagons and octagons result in two distinct forms of OPG, named OPG-*L* (also known as popgraphene) and OPG-*Z*. Fig. 1 illustrates the structural configurations of a 5 × 2 supercell of OPG-*L* and a 3 × 4 supercell of OPG-*Z*, each containing 12 carbon atoms in their minimal cells. Notably, OPG-*L* features octagons in a straight line, whereas those in OPG-*Z* adopt a zigzag pattern. OPG-*L* exhibits varying C-C bond lengths from 1.40 Å to 1.45 Å, with optimized lattice constants of a = 3.68 Å and b = 9.11 Å. In contrast, OPG-*Z* displays C-C bond lengths ranging from 1.39 Å to 1.48 Å and lattice constants of a = 6.90 Å and b = 4.87 Å. These lattice parameters are consistent with previously reported results[1]. The distinct structural details of OPG-*L* and OPG-*Z* elucidate the unique morphological characteristics and bonding behaviors, providing essential insights into their properties and potential applications in diverse fields such as nanotechnology and materials science.

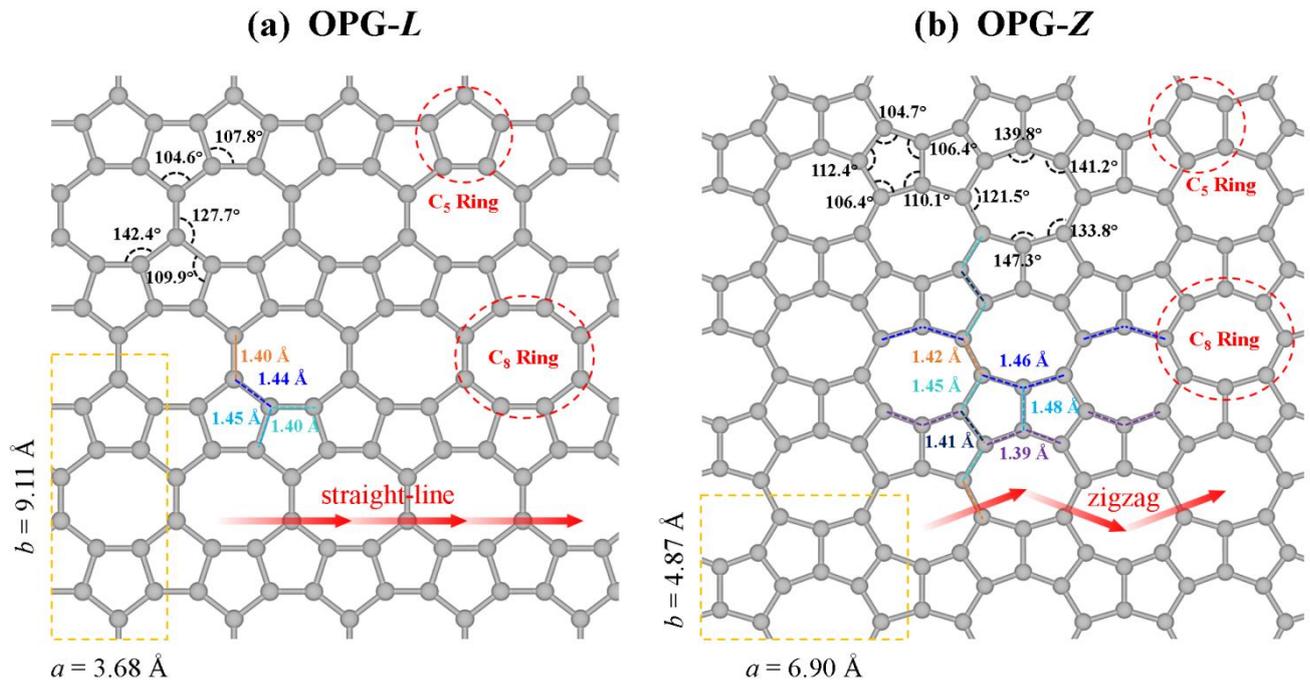

**Fig. 1** Top views of the optimized atomic structure of the (a) OPG-*L* and (b) OPG-*Z*. The yellow dashed rectangle represents the unit cell of OPG-*L* (a = 3.68 Å, b = 9.11 Å) and OPG-*Z* (a = 6.90 Å, b = 4.87 Å). The C-C bonds between adjacent carbon rings are marked by dashed lines with different colors.

*2.2 Computational details*

    All calculations were performed using the Vienna ab initio Simulation Package (VASP)[33-34], based on density functional theory (DFT) and employing the projector-augmented wave (PAW) method[35]. The exchange-correlation energy was described by the generalized gradient approximation (GGA) with the Perdew-Burke-Ernzerhof (PBE) exchange functional[36], and the van der Waals corrections were applied as parametrized in the semiempirical DFT-D3 method. The energy cutoff of the plane wave basis was set to 500 eV. The convergence accuracy for energy was $10^{-5}$ eV within the calculations, and force on each atom was less than 0.02 eV/Å. Spin-polarization was considered in the process of adsorption and dissociation of the oxygen molecule. The Gamma-centered k-point meshes in the Brillouin zone were used for Crystal Orbital Hamilton Populations (COHP) and Bader charge calculations (3 × 3 × 1), structure relaxation and transition states searches (2 × 2 × 1), and molecular dynamic (MD) simulations based on machine learning force field (MLFF). A vacuum region exceeding 12 Å was maintained in all simulations to prevent interactions between periodic images. The 5 × 2 supercell of OPG-*L* and 3 × 4 supercell of OPG-*Z* contains 120 and 144 carbon atoms, respectively. The climbing image nudged elastic band (CI-NEB) method was employed to search the minimum energy path (MEP)[37]. The pymatgen, VASPKIT, and VESTA software packages were employed for data processing and graphics production.

    Recently, the machine-learned force field (MLFF) has emerged as an alternative method for conducting molecular dynamics simulations, significantly extending the length and time scale of MD simulations while maintaining first-principles accuracy[38-39]. It has been utilized to predict various phenomena such as phase transitions[40-41], melting points[42], lattice thermal conductivities[43], and chemical potentials[44]. The essence of the training strategy is the dynamic construction of the MLFF

during MD simulations. At each MD step, the MLFF serves as a criterion to assess the necessity of executing a first-principles calculation and introducing a new structure to the dataset. When further first-principles calculations are deemed unnecessary, the atomic positions and velocities are updated accordingly employing MLFF predictions. We conducted machine-learning-based molecular dynamics (MLMD) simulations using the canonical ensemble (NVT) with the MLFF. To construct the machine-learned interatomic potential, we created 37 periodic structures for OPG-*L* and 28 for OPG-*Z*, each within the respective OPG-*L*/*Z* supercell containing 1-3 randomly positioned oxygen atoms. For each configuration, a 10 ps preheat process from 10 K to 300 K and a 10 ps exploration process with a 1 fs time step were conducted to generate numerous local reference configurations. Ultimately, 3543 and 2663 structures were automatically incorporated into the training sets of OPG-*L* and OPG-*Z*, respectively.

## 3. Results and discussions

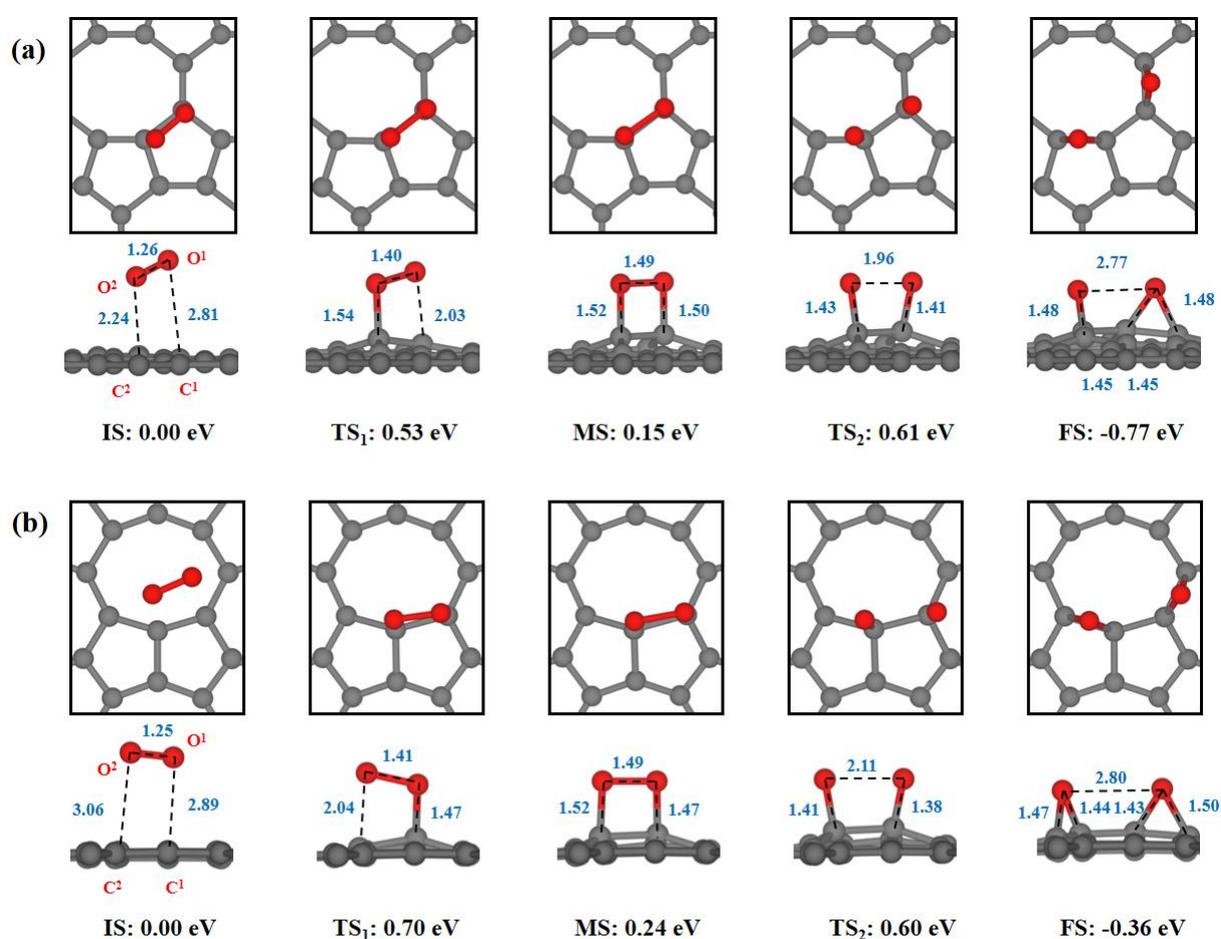

**Fig. 2** Top and side views of the adsorption and dissociation of oxygen molecule on (a) OPG-*L* and (b) OPG-*Z*. The blue and black numbers denote the bond lengths (in Å) and the energy value, respectively. The energy value of the IS is shifted to 0.00 eV for easy comparison both in (a) and (b). Notations: initial state (IS), transition states (TS), and final state (FS).

To delineate the oxidation mechanisms of OPG-*L* and OPG-*Z*, we employed the CI-NEB method to simulate the adsorption and dissociation of oxygen at their respective interfaces. Fig. 2(a) schematically illustrates the adsorption and reaction pathways of an oxygen molecule on the surface of OPG-*L*. Initially, the oxygen molecules are localized in the form of physisorption, positioned approximately 2.2 Å above the OPG-*L* surface, indicating weak van der Waals interactions. As the oxygen molecules approach the surface, their interactions induce significant structural transformations. Specifically, one of the oxygen atoms first forms a C-O bond with a carbon atom that is not aligned with the substrate surface. At this stage, another oxygen atom remains linked to the first oxygen atom through the O-O bond, yet it has not formed a chemical bond with the surface, resulting in an intermediate, partial chemisorption state. The critical transition occurs at transition state $TS_1$, where the C-O bond extends to 2.03 Å, indicating the onset of new chemical bond formation between the oxygen and carbon atoms. The transition state has an energy of 0.53 eV, higher than the initial state's 0.00 eV, representing an activation energy barrier in the reaction pathway. Subsequently, the unbonded oxygen atom forms a second C-O bond with an adjacent carbon atom, resulting in the breaking of the O-O bond. This process occurs at state $TS_2$, where the two oxygen atoms form epoxy groups on the carbon-carbon bonds shared by the $C_5$-$C_8$ ring and $C_8$-$C_8$ ring, respectively. Ultimately, these reactions lead to the chemisorption of the oxygen molecule on the OPG-*L* surface in a cyclized configuration, with an energy of -0.77 eV, lower than that of the initial state, indicating an exothermic adsorption process. Although the adsorption and dissociation processes on both OPG-*L* and OPG-*Z* surfaces are largely analogous, subtle distinctions exist. It is notable that the first transition state ($TS_1$) on OPG-*Z* demands an additional 0.17 eV compared to OPG-*L*, indicating a more substantial activation energy barrier. The subsequent critical transition at $TS_2$ on OPG-Z, where the second C-O bond is formed and the O-O bond is cleaved, is relatively favored by a decrease in the energy barrier by 0.1 eV. This implies that while the initial C-O bond formation is more energy-intensive on OPG-Z, the breaking of the O-O bond and the formation of epoxy groups occur with a lower energy penalty, signifying that the energy costs associated with the formation of chemical bonds can vary significantly across seemingly analogous surfaces.

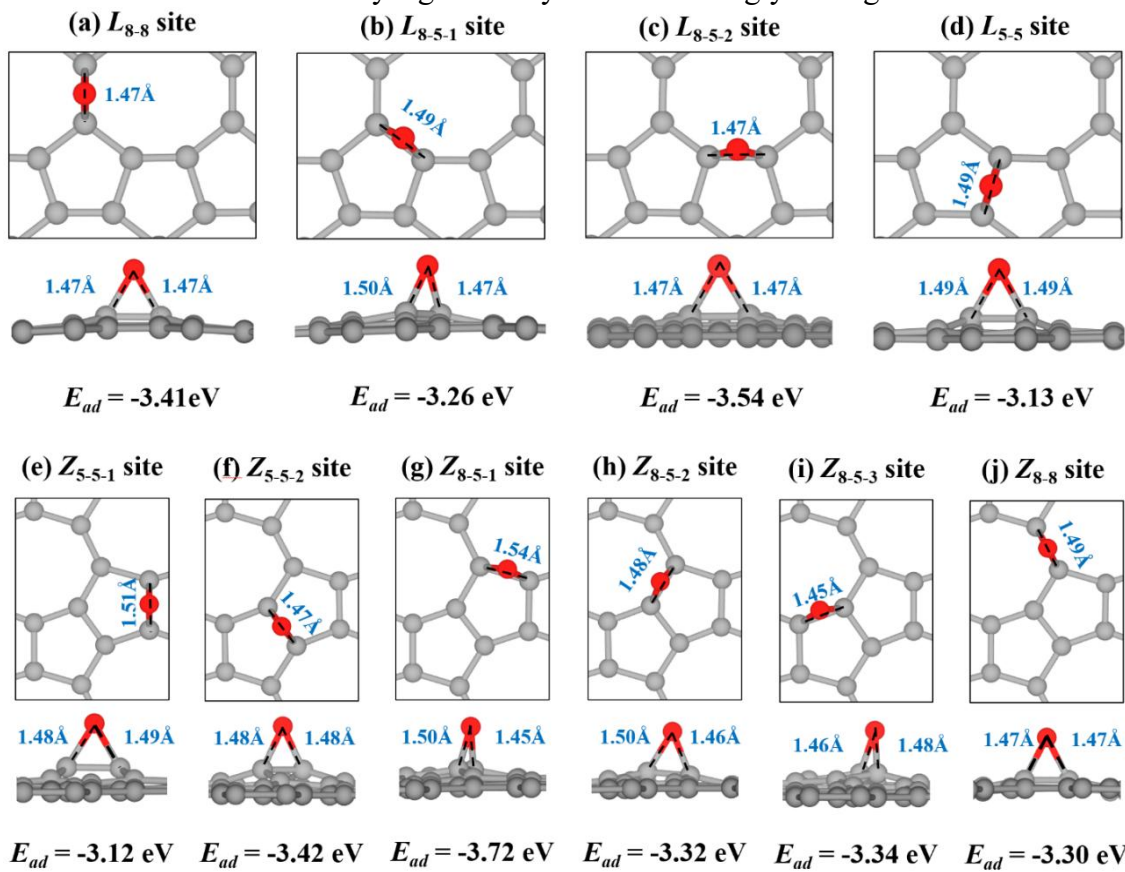

**Fig. 3** Top and side views of the optimized structures of an oxygen atom chemisorbed on different carbon sites on the OPG-*L* (a)-(d) and OPG-*Z* (e)-(j) monolayer: shared C-C bond between the (a) adjacent two $C_8$ rings on OPG-*L*

($L_{8-8}$ site), (b) adjacent $C_5$-$C_8$ rings ($L_{8-5-1}$ site), (c) another adjacent $C_5$-$C_8$ rings ($L_{8-5-2}$ site) and (d) adjacent $C_5$-$C_5$ ($L_{5-5}$ site). Shared C-C bond between the (e) adjacent two $C_5$ rings on OPG-$Z$ ($Z_{5-5-1}$ site), (f) another adjacent two $C_5$ rings ($Z_{5-5-2}$ site), (g) adjacent $C_5$-$C_8$ rings ($Z_{8-5-1}$ site), (h) another adjacent $C_5$-$C_8$ rings ($Z_{8-5-2}$ site), (i) adjacent $C_5$-$C_8$ rings ($Z_{8-5-3}$ site), (j) adjacent two $C_8$ rings ($Z_{8-8}$ site). The adsorption energy is denoted by $E_{ad}$ and the blue numbers denote the bond lengths.

To investigate the physical mechanism underlying oxygen atom adsorption, we present the optimized chemisorption configurations on the OPG-$L$ and OPG-$Z$ monolayers and examine their thermodynamic stability. As illustrated in Fig. 3(a-d), four distinct chemisorption configurations of an isolated oxygen atom on the OPG-$L$ monolayer are explored, including the epoxy group on shared C-C bond between adjacent $C_8$-$C_8$ ($L_{8-8}$ site), $C_8$-$C_5$ ($L_{8-5-1}$ and $L_{8-5-2}$ sites), and $C_5$-$C_5$ ($L_{5-5}$ site). These configurations correspond to four different non-equivalent carbon sites. Meanwhile, as depicted in Fig. 3(e-j), we identify six possible chemisorption configurations on the OPG-$Z$ monolayer, including epoxy on the shared C-C bond between adjacent $C_5$-$C_5$ ($Z_{5-5-1}$ and $Z_{5-5-2}$ sites), $C_8$-$C_5$ ($Z_{8-5-1}$, $Z_{8-5-2}$, and $Z_{8-5-3}$ sites) and $C_8$-$C_8$ ($Z_{8-8}$ site). The stability of oxygen molecule adsorption can be quantitatively described by the adsorption energy $E_{ad}$, defined as $E_{ad} = E_{adsorbate@OPG-L/Z} - E_{OPG-L/Z} - E_{adsorbate}$, where $E_{adsorbate@OPG-L/Z}$, $E_{OPG-L/Z}$ and $E_{adsorbate}$ represent the energies of the OPG-$L/Z$ monolayer with and without the adsorbate, and an isolated adsorbate, respectively. The corresponding adsorption energies are detailed in Fig. 3. The $E_{ad}$ for the epoxy group on the OPG-$L$ indicates that the lowest energy of -3.54 eV occurs at the $L_{8-5-2}$ site, while highest energy of -3.13 eV is at the $L_{5-5}$ site. This suggests that the oxygen atom preferably chemisorbs at the $L_{8-5-2}$ site, as evidenced by the relatively shorter bond length. Additionally, we identify the $Z_{8-5-1}$ site on the OPG-$Z$ as the energetically favorable configuration, featuring the lowest adsorption energy of -3.72 eV. Conversely, the adsorption energy of the epoxy group attached to the carbon atom on $Z_{5-5-1}$ site is -3.12 eV. The lower $E_{ad}$ values at the $Z_{8-5-1}$ site indicates a more stable configuration.

In an effort to quantitatively assess the interactions within C-O bonds, we conducted an integrated-crystal orbital Hamilton population (ICOHP) analysis, integrating the band states up to the highest occupied energy level. This sophisticated approach unveiled variations in the strength of C-O bonding across different sites within the material. As depicted in Fig. 4 (a), the ICOHP values at the $L_{8-5-2}$ site are notably strong, measuring -7.99 eV and -8.11 eV, indicative of robust bonding. In stark contrast, the ICOHP values for the C-O bonds in Fig. 4 (b) are -7.81 eV and -7.86 eV, respectively, which are less negative and thus suggest a considerably weaker bond at the $L_{5-5}$ site compared to the $L_{8-5-2}$ site. The COHP analysis at the $L_{5-5}$ site also exhibits a pronounced antibonding contribution, reinforcing the conclusion of a weaker C-O interaction at this site. In Fig. 4 (c-d), the ICOHP values for OPG-$Z_{5-5-1}$ and OPG-$Z_{8-5-1}$ are -8.23 eV, -7.83 eV, -7.61 eV, and -8.87 eV, respectively. A nuanced comparison of these values reveals that while the ICOHP$_1$ value for OPG-$Z_{5-5-1}$ is marginally higher than that for OPG-$Z_{8-5-1}$, the ICOHP$_2$ for OPG-$Z_{8-5-1}$ markedly surpasses its counterpart. This discrepancy is pivotal, as it correlates with the higher adsorption energy of oxygen atoms at the $Z_{8-5-1}$ site compared to the $Z_{5-5-1}$ site. Furthermore, the COHP analysis at the $Z_{5-5-1}$ site indicates a heightened antibonding contribution, emphasizing the diminished bond strength between carbon and oxygen at this site.

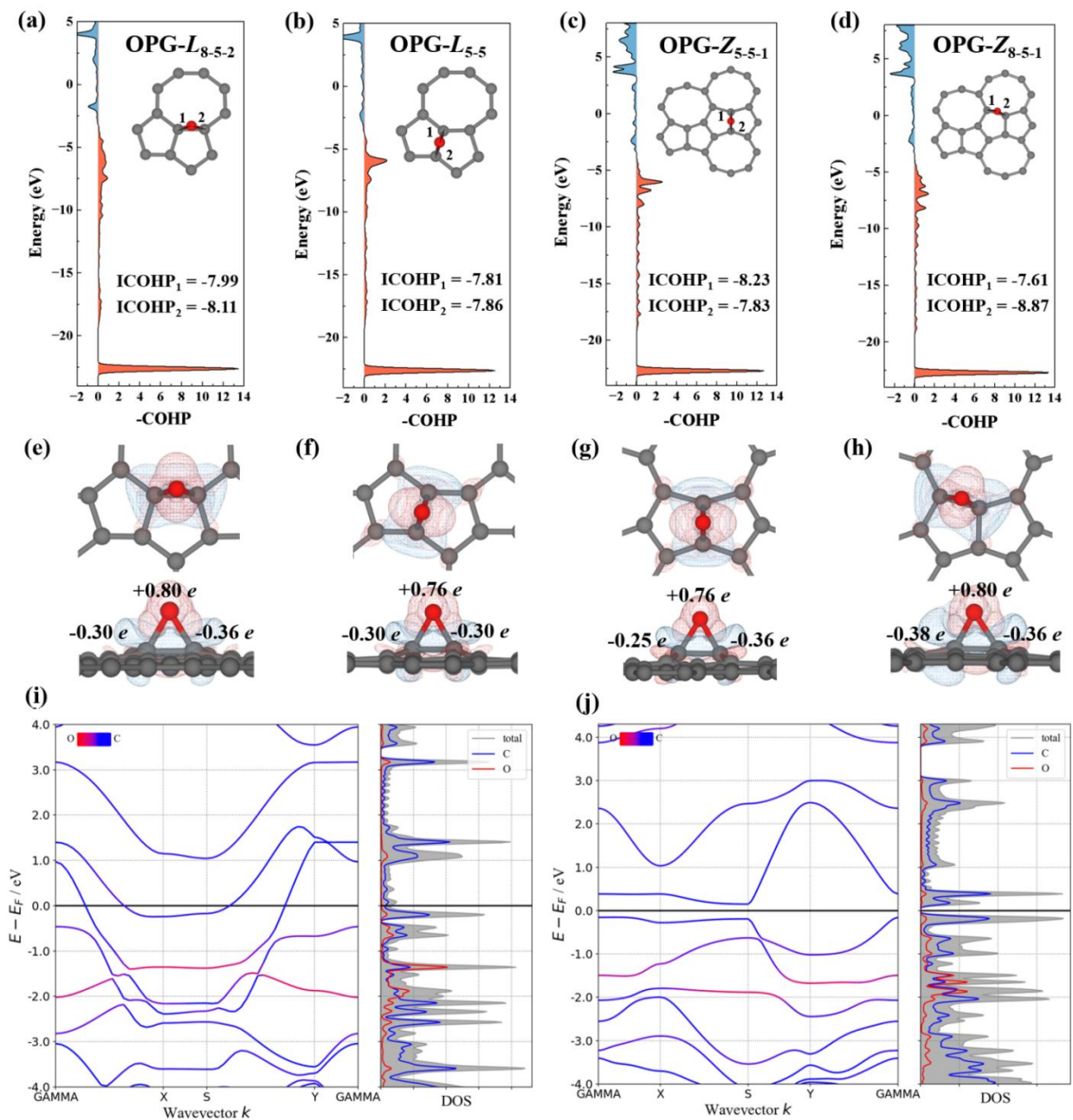

**Fig. 4** (a-d) The crystal orbital Hamilton populations (COHP) profiles for the most and least energetically favorable oxygen adsorption sites on OPG-$L$ and OPG-$Z$. (e-h) The charge density difference and Bader charge of the epoxy with the four structures. Details of the electronic band structures and density of states (DOS) are shown for the epoxide structure at $L_{8-5-2}$ site (i) and at $Z_{8-5-1}$ site (j), respectively.

To further elucidate the physical mechanisms governing the stable adsorption of oxygen atoms on OPG-$L$ and OPG-$Z$ sites, Fig. 4(e-h) presents a comparative analysis of the charge density difference and Bader charges for four distinct product configurations. The computational findings indicate that the epoxide structure in Fig. 4 (e) at the OPG-$L_{8-5-2}$ site exhibits an oxygen atom with an elevated electron count of +0.80 e, acquired from adjacent carbon atoms, surpassing the count of +0.76 e observed in the structure of Fig. 4 (f). Similarly, at the OPG-$Z$ site, the epoxide structure in Fig. 4 (g) demonstrates a higher electron density with the oxygen atom, recorded at +0.80 e, compared to +0.76 e in the structure depicted in Fig. 4 (h). These variations are attributed to the high electronegativity of the oxygen atom, enhancing its capacity to attract electrons. The enhanced electron density not only strengthens the C-O bond but also significantly increases the adsorption energy of oxygen atoms at the OPG-$L_{8-5-2}$ and OPG-$Z_{8-5-1}$ sites. The increase of adsorption energy directly correlates with the enhanced thermodynamic stability of the epoxy structure. In addition, Fig. 4 (i-j) show the electronic band structures and density of states (DOS) of the the epoxide structure at the OPG-$L_{8-5-2}$ site and OPG-$Z_{8-5-1}$ site. Compared to the pristine OPG-$L$ displayed in Fig. S1, we observe that the adsorption of oxygen does not alter the metallic properties of OPG-$L$. By contrast, the introduction of oxygen can create a band gap at the OPG-$Z_{8-5-1}$ site, thereby effecting a transformation of OPG-$Z$ from a gapless metallic state to a semiconducting phase.

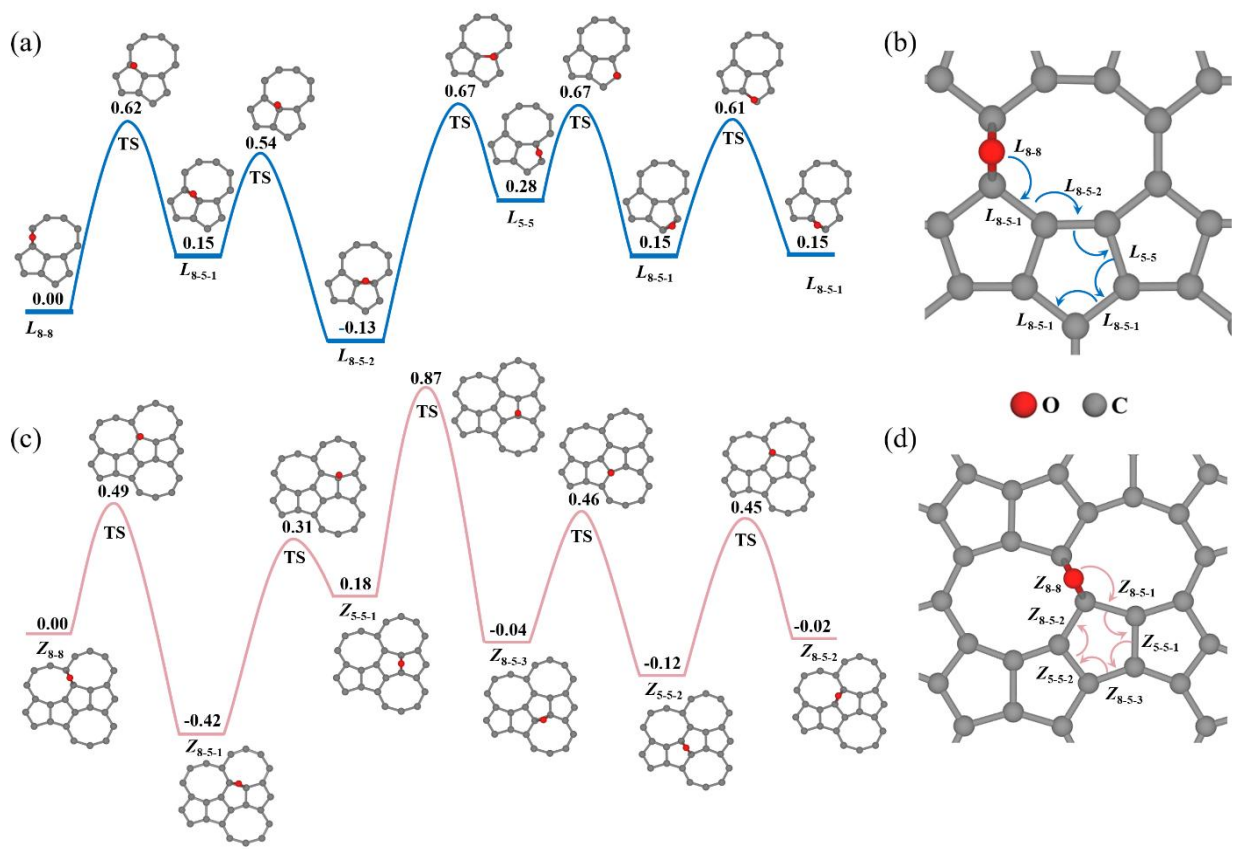

**Fig. 5** The oxygen migration pathways and state configurations on the surface of OPG-$L$ (a, b) and OPG-$Z$ (c, d). The energy value of $L_{8-8}$ and $Z_{8-8}$ is shifted to 0 eV. The non-equivalent carbon sites are labeled by the same notations used in Fig. 2. The direction of the arrow indicates the possible pathways of the oxygen migration. Extra migration pathways on the OPG-$Z$ are shown in the Fig. S2.

The diverse chemisorption configurations result in varying oxygen migration pathways, with noted limitations in oxygen migration specifically along the basal plane of OPG. Fig. 5 provides a comprehensive visualization of oxygen migration pathways and state configurations on the surfaces of the OPG-$L$ and OPG-$Z$ monolayers. Five elementary oxygen migration pathways interconnect four distinct oxygen chemisorption sites on OPG-$L$, as delineated in Fig. 5(a-b). Oxygen migration on OPG occurs via the breaking or reforming of C-O bonds. Specifically, on OPG-$L$, oxygen atoms can migrate from the $L_{8-8}$ site to the $L_{8-5-1}$ site by breaking the C-O bond, resulting in a dangling oxygen configuration as a transient state, followed by C-O bond formation with another carbon atom, yielding an epoxy group at the $L_{8-5-1}$ site. In this process, an energy barrier of 0.62 eV is required to break the C-O bond, thereby facilitating the oxygen migration between $L_{8-8}$ site and $L_{8-5-1}$ site.

Subsequently, the oxygen atom overcomes a relatively low energy barrier of 0.39 eV to reach $L_{8\text{-}5\text{-}2}$ site, which is more stable. It is important to note that the energy barrier between the $L_{8\text{-}5\text{-}2}$ and $L_{5\text{-}5}$ sites is significantly higher at 0.80 eV, suggesting greater stability at the $L_{8\text{-}5\text{-}2}$ site. This observation aligns with previous findings depicted in Fig. 3 and 4. At $L_{5\text{-}5}$ site, oxygen can migrate to $L_{8\text{-}5\text{-}1}$ site with greater ease through the breaking and reforming of the C-O bond, which is associated with a lower barrier of 0.39 eV. Moreover, the energy barrier between two equivalent $L_{8\text{-}5\text{-}1}$ sites is 0.46 eV. Overall, the energy barriers for oxygen migration pathways on the OPG-$L$ surface range from 0.39 eV to 0.80 eV, indicating a spectrum of stability across different oxygen adsorption configurations.

A similar oxygen migration mechanism exists on the OPG-$Z$ monolayer as that observed on OPG-$L$ exists. This mechanism involves the migration of oxygen atoms via the breaking and reforming of C-O bonds. However, due to the zigzag arrangement of the carbon rings on the OPG-$Z$, the migration pathways and corresponding energy barriers differ from those on the OPG-$L$. The energy barriers for oxygen migration on OPG-$Z$ typically range from 0.45 eV to 0.91 eV. Notably, the migration from $Z_{5\text{-}5\text{-}1}$ to $Z_{8\text{-}5\text{-}1}$ features an energy barrier significantly lower at 0.13 eV. While this low barrier suggests a higher likelihood of oxygen moving from $Z_{5\text{-}5\text{-}1}$ to $Z_{8\text{-}5\text{-}1}$, it also implies that oxygen will remain stable once it reaches $Z_{8\text{-}5\text{-}1}$. This indicates restricted migration of oxygen along the basal plane of OPG-$Z$. Furthermore, a high energy barrier of 0.73 eV from $Z_{8\text{-}5\text{-}1}$ site back to $Z_{5\text{-}5\text{-}1}$ site also corroborates previous findings about energetically favorable configuration on OPG-$Z$. Our DFT calculations indicate that oxygen migrations on both the OPG-$L$ and OPG-$Z$ monolayers are not spontaneous and exhibit low migration rates due to substantial energy barriers.

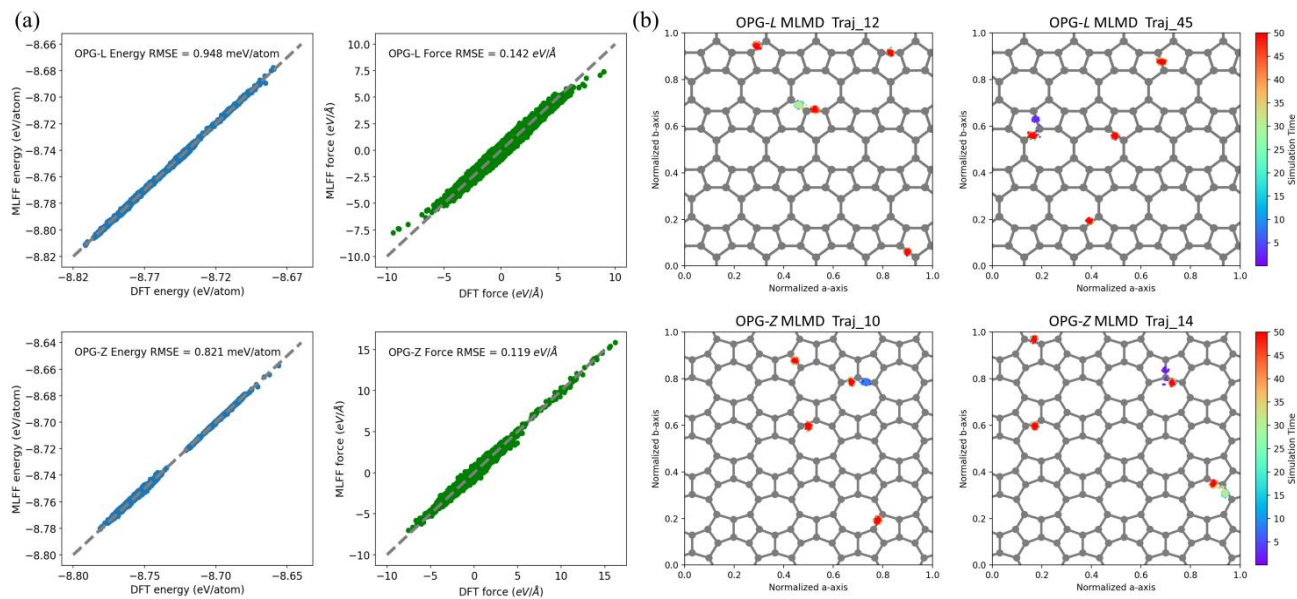

**Fig. 6** Machine learning modeling and MLMD simulations of oxygen migration on OPG-$L$ and OPG-$Z$ monolayers. (a) Energy and force as calculated from the MLFFs compared with the training data from DFT calculations. (b) MLMD simulations trajectories of oxygen migration on OPG-$L$ (above) and OPG-$Z$ (below). The colored dots denote the positions of 4 oxygen atoms, and the colors denote the simulation time. The static OPG monolayer is used as the background for clarity.

The limited oxygen migration on OPG-$L$ and OPG-$Z$ monolayers can be further corroborated by MLMD simulations. Prior to this, to confirm the accuracy of the MLFFs, energies and forces were calculated and compared with DFT results for the training set, which consists of 3543 OPG-$L$ and 2663 OPG-$Z$ structures, as depicted in Fig. 6(a). For the OPG-$L$ system, our analysis indicates a root mean square error (RMSE) of 0.948 meV/atom for energies and 0.142 eV/Å for forces, highlighting the precision of our MLFFs. Similarly, for the OPG-*Z* system, the RMSE values are 0.821 meV/atom for energies and 0.119 eV/Å for forces, further corroborating the reliability of our machine learning models and MLMD simulations[40, 45]. These results demonstrate the remarkable accuracy of our computational framework, showcasing its potential to advance predictive modeling in materials science with profound implications across diverse applications. The MLMD trajectories of oxygen migration are presented for both the OPG-*L* and OPG-*Z* monolayers over a 50 ps timescale, as shown in Fig. 6(b). To thoroughly investigate the migration behavior, 50 random samples were generated for each OPG system, with four oxygen atoms randomly introduced into each sample. These samples were subjected to 50 ps MD simulations, with four representative trajectories selected for detailed analysis. In the OPG-*L* system, the trajectories reveal that oxygen atoms migrate from the $L_{8\text{-}5\text{-}1}$ site to $L_{8\text{-}5\text{-}2}$ site and from the $L_{5\text{-}5}$ site to $L_{8\text{-}5\text{-}1}$ site during the simulation period. This observation is consistent with our CI-NEB calculations, which indicate an energy barrier of 0.39 eV for such movements, suggesting that higher energy barriers inhibit spontaneous migration across the OPG-*L* monolayer. On the other hand, in the OPG-*Z* system, a migration process from $Z_{5\text{-}5\text{-}1}$ to $Z_{8\text{-}5\text{-}1}$ is observed between about 15ps-17ps, which is reasonable given the low energy barrier of 0.13 eV for this specific pathway. Furthermore, two additional migration processes involving movements from $Z_{8\text{-}5\text{-}1}$ to $Z_{8\text{-}5\text{-}1}$ and from $Z_{8\text{-}5\text{-}2}$ to $Z_{8\text{-}5\text{-}3}$ are observed, which exhibit energy barriers of 0.46 eV and 0.45 eV, respectively. These findings are corroborated by the CI-NEB calculations detailed in Fig. S3. In the remaining 48 samples, within the simulated timescales on both OPG-*L* and OPG-*Z* monolayers, no additional pathways for oxygen migration are detected, aligning with our previous observations. This consistency reinforces the validity of the identified oxygen migration patterns on the OPG interfaces. Overall, the MLMD simulations provide valuable insights into the dynamics of oxygen migration on the OPG-*L* and OPG-*Z* monolayers, complementing the DFT calculations and providing a comprehensive understanding of these dynamic properties.

## 4. Conclusion

In summary, our systematic computational study reveals the complex oxidation mechanisms of OPG and the fundamental interactions between OPG and oxygen by employing the combination of DFT calculations and MLMD simulations. The energy barriers that accompany the physisorption, chemisorption and dissociation processes of oxygen molecules, as well as regulate the oxygen migration along the basal plane of OPG shows the dynamic behaviors of oxygen atoms at the OPG interface. In addition, we identify significant variations in the energy barriers for oxygen migration across different chemisorption sites on the OPG monolayer, with values ranging from 0.39 eV to 0.91 eV, underscoring the material's unique response to an oxygen-rich environment. These findings contribute to a deeper understanding of the dynamic behaviors of oxygen on this distinctive octa-penta topological structure, which positions it as a promising candidate for applications requiring surface stability and oxygen resistance. Furthermore, our study lays a solid foundation for subsequent experimental explorations and further broadens the horizon for exploration into interactions between OPG and other reactive species.

**Author Contributions**
Z. Yan conceived and designed the research; Y. Tu, Z. Yan, and Z. Zhang guided the research; C. Zhou and B. Situ performed the simulations; C. Zhou, R. Huo, Z. Yan and Z. Zhang wrote the paper;

All the authors participated in discussions of the research.

**Conflicts of interest**
There are no conflicts to declare.

**Acknowledgement**
This work was funded by the National Natural Science Foundation of China (Nos. 12347169), the Natural Science Foundation of Jiangsu Province (No. BK20240892), the Outstanding Doctor Program of Yangzhou City "Lv Yang Jin Feng" (YZLYJFJH2022YXBS087), the China Postdoctoral Science Foundation (No. 2023M732998) and Jiangsu Students' Innovation and Entrepreneurship Training Program (No. 202311117038Z). Numerical computations were performed on Hefei advanced computing center.